\documentclass[showpacs,aps,epsfig,nofootinbib]{revtex4}
\usepackage[T2A]{fontenc}

\begin{document}
\title{Non-adiabatic perturbations in multi-component perfect fluids.}

\author{N.A. Koshelev}
\email{koshna71@inbox.ru}

\affiliation{Ulyanovsk State University, Leo Tolstoy str 42,
432970, Russia}

\date{\today}

\begin{abstract}

The evolution of non-adiabatic perturbations in models with
multiple coupled perfect fluids with non-adiabatic sound speed is
considered. Instead of splitting the entropy perturbation into
relative and intrinsic parts, we introduce a set of symmetric
quantities, which also govern the non-adiabatic pressure
perturbation in models with energy transfer. We write the gauge
invariant equations for the variables that determine on a large
scale the non-adiabatic pressure perturbation and the rate of
changes of the comoving curvature perturbation. The analysis of
evolution of the non-adiabatic pressure perturbation has been made
for several particular models.

\end{abstract}


\maketitle

\section{Introduction.}

Multi-component models play an important role in modern cosmology.
They are required to study a large scale structure formation
\cite{Kodama_Sasaki,LL93} and are used in multi-field inflationary
models \cite{Linde_1985,Starobinsky_1985,KLF_1985}. Multi-fluid
models are also essential in the context of preheating
\cite{KLS_preheat94,KLS_preheat97}. It is well known that in
cosmological models with multiple fractions the growing mode
solution of the curvature perturbation on comoving hypersurfaces
${\cal R}$ remains constant on a large scale in the absence of any
entropy perturbations \cite{Kodama_Sasaki,LL93,LSWL}. This
quantity is valuable to cosmological applications, because, for
example, it allows one to relate the perturbations generated at a
stage of inflation with the primordial perturbations in the early
radiation dominated era. In the presence of the non-adiabatic
pressure perturbations the comoving curvature perturbation ${\cal
R}$ always changes with time. Thus, the non-adiabatic
perturbations may affect the observable power spectrum of the
adiabatic perturbations.

A scalar field and a cosmological fluid seem very different,  but
the distinctions are not fundamental. Any barotropic perfect fluid
can be represented as a scalar field with a nontrivial kinetic
part \cite{0806.1016}. On the other hand, in the linear
perturbation theory the scalar field with a self-coupling
potential can be treated as a perfect fluid with sound speed
$c_s^2\equiv \delta P/\delta\rho$ (computed in the fluid rest
frame) different from adiabatic sound speed $c_{s(ad)}^2$
\cite{Gordon_Hu}. For this reason, an effective fluid description
is often used. Under this approach, a scalar field can be fully
described by specifying its sound speed and the equation of state,
and by phenomenologically introducing  an energy-momentum transfer
between the scalar field and other fluids. In slightly different
and more general formalism, any system of $ N $ coupled canonical
scalar fields can be modelled by $ N $ "kinetic" fluids with a
stiff equation of state interacting with one "potential" fluid
with a vacuum equation of state \cite{Malik_Wands2004}.

A valuable class of multi-component models is formed by the
systems containing the coupled scalar fields and the barotropic
fluids. These are, among others, the cosmological models with
coupled cold dark matter and dark energy, considered as a
canonical scalar field usually called quintessence. For these
models there has been found the possibility of existence of large
scale instabilities at the radiation dominated stage, caused by
the fast growth of non-adiabatic perturbations \cite{VMM_0804}.
Such early time instabilities have received much attention in a
variety of scenarios \cite{0807.3471,0901.1611,0901.3272}.

Although non-adiabatic perturbations were investigated long ago,
there still remain some unresolved issues.  Currently, the
definitions of the relative entropy perturbation for scalar fields
\cite{PSt92,PSt94} (note that the authors of Refs.
\cite{GWBM,Hwang_Noh} used adiabatic and entropy field combinations
instead) and of general perfect fluids with intrinsic entropy
perturbations \cite{Malik_Wands2004,Malik_Wands_Ungarelli} are
different. Thought Ref. \cite{Malik_Wands2004} gives a unified
definition of relative entropy that is appropriate both for
barotropic fluids and for scalar fields, its use leads to some
theoretical difficulties. In particular, even in the simplest single
field models this method assumes, in general, a nonzero relative
entropy between the "kinetic" and "potential" fluids, which seems
somewhat artificial. In the phenomenological approach there are also
some problems with the description of the interaction of these
auxiliary fluids and others. It is also desirable to study the
adiabatic condition in more detail. For example, in the recent paper
\cite{1004.5509} an analytical attractor solution for dark energy
perturbations in the synchronous gauge at a constant dark energy
equation of state parameter was obtained. For this solution, the
usually imposed generalized adiabatic condition
\cite{GWBM,Hwang_Noh} does not take place.

The aim of this work is to, at least partially, fill the gaps. The
paper is organized as follows. In Section \ref{basic} we briefly
review the perturbed Einstein and continuity equations for a
general multi-fluid model with energy exchange. Assuming that the
total energy density can be represented as a sum of energy
densities of some perfect fluids, we write, in Section
\ref{entropy}, an evolution equation for the comoving curvature
perturbation ${\cal R}$ and an expression for the non-adiabatic
pressure perturbation $P_{nad}$. Instead of using of relative
entropy perturbations, we introduce a set of symmetric quantities
$\tilde{S}_{IJ}$ that determine the $P_{nad}$ in models of
considered type. In Section \ref{entropy} we also derive the gauge
invariant equations for variables that describe the evolution of
non-adiabatic density and pressure perturbations. Some
applications of the proposed formalism are discussed in Section
\ref{appl}. In particular, the case of two minimally coupled
fluids and applications to cosmological models with coupled dark
matter and quintessence dark energy is studied in detail. We
conclude the paper in Section \ref{concl}.

\section{Basic equations.}
\label{basic}

Let us consider scalar type perturbations at a FRW background. The
general line element for scalar perturbations is
\begin{equation}
\label{metric}ds^2 = a^2(\tau)\left\{-(1+2\phi) d\tau^2
+2B_{,i}d\tau dx^i+[(1 - 2\psi) \delta _{ij} +2E_{,ij}]dx^idx^j
\right\},
\end{equation}
where four scalar $\phi$, $\psi$, $B$, $E$ are the first-order
quantities and $\tau$ is the conformal time.

A perfect  $I$-fluid with density $\rho_I
=\bar{\rho}_{I}+\delta\rho_{I} $, pressure $P_I=\bar{P}_I+\delta
P_I$, four-velocity $ u_{I}^\mu = dx_{I}^\mu/ds $ and vanishing
anisotropic stress is defined to have an energy-momentum tensor of
the form
\begin{equation}
\label{tensor}T_{I\nu} ^{\mu}  = \left(\rho_I + P_I
\right)u_I^{\mu} u_{I\nu} + P_I\delta _{~\nu}^{\mu}  ,
\end{equation}
The perturbed four-velocity  can be written to the first-order
terms as
\begin{equation}
\label{velocity}u_I^{\mu} = \frac{1}{a} \left[ \left( 1-\phi
\right), ~v_{I}^{,i} \right] , ~~~~~~ u_{I\nu} = a \left[
-(1+\phi),~v_{I,i}+B_{,i} \right].
\end{equation}
Here, we introduce the velocity potential $ v_I $, since the fluid
flow is irrotational for scalar perturbations.

Throughout this paper, we will work in the Fourier space. The
first-order perturbed Einstein equations yield
\begin{eqnarray}
\label{Ein1}3{\cal H}\left( {\psi ' + {\cal H}\phi} \right)  +k
^{2}\psi + k ^{2}{\cal H}\left({E' - B} \right) &=& -
4\pi G a^{2}\delta \rho, \\
\label{Ein2}\psi ' + {\cal H}\phi &=&  -4\pi G
a^{2}\left( \bar{\rho} + \bar{P} \right)(B+v_I),\\
\psi'' + 2{\cal H}\psi ' + {\cal H}\phi ' + \left( {2{\cal H}' +
{\cal H}^{2}} \right)\phi &=&4\pi G a^{2} \delta P ,\\
\left( {E' - B} \right)' +  2{\cal H}\left( {E' - B} \right) -
\phi + \psi &=& 0,
\end{eqnarray}
where $k$ is a comoving wave number,  the prime denotes the
differentiation with respect to the conformal time $\tau$, and
${\cal H}=a'/a$.

In the general case of $N$ coupled fluids divergence of the
energy-momentum tensor gives
\begin{equation}
\label{div_eq}T^{\mu\nu}_{I;\nu}=Q^{\mu}_{I},
\end{equation}
The four-vectors $Q^{\mu}_{A}$ are related by the constraint
\cite{Kodama_Sasaki}
\begin{equation}
\sum_{I=1}^{N} Q_I^\mu =0,
\end{equation}
which results from the conservation law of the total
energy-momentum tensor. For convenience, one can decompose these
4-vectors into two parts
\begin{equation}
Q_I^\mu =Q_I u^\mu +F_I^\mu, ~~~~ Q_I=\bar{Q}_I+ \delta Q_I,
~~~~u_\mu F_I^\mu =0.
\end{equation}
Here $u^\mu$ is the overall four-velocity, $Q_I$ is the energy
density transfer rate and $F_I^\mu$ is the momentum density
transfer rate of $I$-fluid in the total matter gauge. We have
$F_I^\mu =a^{-1}(0,f_I^{~,i})$ up to first-order, where $f_I$ is a
momentum transfer potential \cite{VMM_0804}.

The continuity equations for coupled perfect fluids can be
obtained by the linearization of conservation equations
(\ref{div_eq}). As a result, one can write the background
\begin{equation}
\bar{\rho}_I'=-3{\cal H}(1+w_I)\bar{\rho}_I+a\bar{Q}_I
\end{equation}
and the perturbed equations \cite{VMM_0804}
\begin{eqnarray}
\label{cont_1}\delta\rho_{I}' + 3{\cal
H}\left(\delta\rho_{I}+\delta P_{I}\right) -  3\left(
\bar{\rho}_{I} + \bar{P}_{I}\right) \psi' -k^2 \left(
\bar{\rho}_{I} + \bar{P}_{I}\right) \left(v_I+E'
\right) &=& a \bar{Q}_I\phi +a\delta Q_I,\\
\label{cont_2}\left[(\bar{\rho}_I\!+\bar{P}_I) (v_I+B) \right]' +
4{\cal H}(\bar{\rho}_I\!+ \bar{P}_I)(v_I+B)+ (\bar{\rho}_I\!+
\bar{P}_I)\phi +\delta P_I &=& a \bar{Q}_I(v+B) +af_I .
\end{eqnarray}

Pressure perturbations can be expressed in terms of density and
velocity perturbations as \cite{VMM_0804}
\begin{equation}
\delta P_I  =c^2_{sA}\delta \rho_I - (c^2_{sI}-c^2_{sI(ad)})
\rho_I' \frac{\theta_I}{k^2}.
\end{equation}
where the velocity perturbation $\theta_I =-k^2(B+v_I)$,
$c_{Is(ad)} ^2 = P_I'/\rho_I'$ is the $I$-fluid adiabatic sound
speed and $c_{sI} ^2 =\delta P_I /\delta \rho_I $ is defined in
the $I$-fluid rest frame. The values of $c_{Is(ad)} ^2 $ and
$c_{sI} ^2$ are equal for the barotropic fluid (i.e. fluid for
which the pressure depends only on the density), but this equality
may not be satisfied in a general case. For example, for a
canonical scalar field one has to set $c_{sI} ^2 =1$
\cite{Gordon_Hu}.

Using the notation
\begin{equation}
q_{I}\equiv \frac{a \bar{Q}_{I}}{3{\cal H} (\bar{\rho}_I +
\bar{p}_I)}
\end{equation}
and the identity
\begin{equation}
\label{constr} w_I'=\left(\frac{\bar{P}_I}{\bar{\rho}_I} \right)'=
3{\cal H} (1+w_I) (1-q_I)\left(w_I - c^2_{sI(ad)} \right) ,
\end{equation}
the equations (\ref{cont_1}),(\ref{cont_2}) can be rewritten as
\begin{eqnarray}
\left(\Delta_I - 3\psi\right)' &+& 3{\cal H} \left(c^2_{sI}
-c^2_{sI(ad)}+ q_I(c^2_{sI(ad)} +1)  \right) \Delta_I \nonumber \\
\label{cont_1a}&& + 9{\cal H}^2 (c^2_{sI} - c^2_{sI(ad)})(1 - q_I
) \frac{ \theta_I}{k^2}  + \theta_I - k^2 \left(E'-B \right)=
3{\cal H} q_I \left( \phi +\frac{\delta Q_I}{\bar{Q}_{I}}\right),\\
\label{cont_2a}\frac{\theta_I'}{k^2} +{\cal H}\frac{ \theta
_I}{k^2} &=& c^2_{sI} \left(\Delta_I +3{\cal H}(1 - q_I )\frac{
\theta _I}{k^2}\right) + \phi + 3{\cal H}q_I
\left(\frac{\theta}{k^2}- \frac{\theta_I}{k^2} \right) -
\frac{a}{\rho_I (1+w_I)}f_I,
\end{eqnarray}
where $\Delta_I=\delta\rho_I/( \bar{\rho}_{I} + \bar{P}_{I})$.

\section{Non-adiabatic perturbations.}
\label{entropy}

The comoving curvature perturbation $ {\cal R} = \psi - {\cal H}
\left( v+B \right)$  at spatially flat background can be
represented in terms of the longitudinal gauge-invariant
quantities \cite{MFB}
\begin{equation}
\Phi \equiv \phi + {\cal H}(B-E') + (B-E')'  ,~~~~~~ \Psi \equiv
\psi - {\cal H} \left( B-E' \right)
\end{equation}
as
\begin{equation}
\label{curvature}{\cal R} = \Psi + \frac{2}{3}\frac{ \Psi' + {\cal
H} \Phi} {{\cal H} (1+w) } .
\end{equation}

The time derivative of equation (\ref{curvature}) gives the well
known equation
\begin{equation}
{\cal R}' = \frac{2}{3}\frac{1}{ {\cal H}\left(1+w \right)}
\left\{ \frac{ \kappa ^2}{2} a^{2} P_{nad} -k^2c^2_{s(ad)} \Psi
\right\},
\end{equation}
where the non-adiabatic part of the pressure perturbation is
defined by
\begin{equation}
P_{nad}=\delta P -c^2_{s(ad)}\delta\rho .
\end{equation}

Substituting the expression $c^2_{s(ad)}=\sum\limits_I
\frac{\bar{\rho}_I'}{ \bar{\rho}'} c^2_{sI(ad)}$ for the overall
adiabatic sound speed, at the arbitrary number of coupled perfect
fluids, one can obtain
\begin{eqnarray}
P_{nad} &=& \sum_I \frac{\bar{\rho}_I'}{\bar{\rho}'}
\left(c^2_{sI}- c^2_{sI(ad)} \right)\epsilon_m + \sum_I
(\bar{\rho}_I+\bar{P}_I)c^2_{sI}
q_I\Delta \nonumber \\
\label{pnad_1}&&+ \sum_{I,J}   \frac{ (\bar{\rho}_I+\bar{P}_I)
(\bar{\rho}_J+\bar{P}_J)}{\bar{\rho}+\bar{P}} \left(
c^2_{sI}\Delta _{IJ}+ 3{\cal H}\left(c^2_{sI}- c^2_{sI (ad)}
\right) (1-q_I) \frac{\theta_{IJ}}{k^2}\right),
\end{eqnarray}
where $\Delta =\delta\rho/( \bar{\rho} + \bar{P})$, $\Delta _{IJ}
=\Delta _{I} -\Delta _{J}$, and the gauge-invariant quantity
\begin{equation}
\label{epsilon_m}\epsilon_m \equiv \delta \rho + 3{\cal H}\left(
\bar{\rho} + \bar{P} \right)\frac{\theta}{k^2}
\end{equation}
is a comoving density perturbation \cite{Bardeen}.

The equation (\ref{pnad_1}) contains quantities $ \Delta_{IJ}$
that are not gauge invariant at non-minimal coupling. Hence,
following Refs. \cite{Malik_Wands2004},
\cite{Malik_Wands_Ungarelli}, \cite{Malik_PhD}, it is convenient
to use the gauge invariant variables
\begin{equation}
\hat{\Delta}_{IJ} = \frac{\delta\rho_I}{(1-q_I) (\bar{\rho}_I
+\bar{P}_I)} -\frac{\delta\rho_J}{(1-q_J) (\bar{\rho}_J +
\bar{P}_J)}
\end{equation}
or
\begin{equation}
\tilde{\Delta} _{IJ} = (1-q_I)(1-q_J)  \hat{\Delta} _{IJ}
\end{equation}
instead. The latter quantities are well defined even at $ q_I = 1
$.

In these new variables, the expression (\ref{pnad_1}) can be
rewritten as
\begin{equation}
\label{pnad_2} P_{nad}= \sum_I \frac{\bar{\rho}_I'}{\bar{\rho}'}
\left( c^2_{sI} - c^2_{sI(ad)} \right)\epsilon_m +
\frac{1}{2}\sum_{I,J} \frac{ (\bar{\rho}_I+\bar{P}_I)
(\bar{\rho}_J+\bar{P}_J)}{\bar{\rho} +\bar{P}} \tilde{S}_{IJ},
\end{equation}
where
\begin{equation}
\label{psevdo}\tilde{S}_{IJ}= \left(c^2_{sI} -c^2_{sJ}\right)
\tilde{\Delta} _{IJ}+ 3{\cal H}\left( \left(c^2_{sI}- c^2_{sI
(ad)} \right) (1-q_I) - \left(c^2_{sJ}- c^2_{sJ(ad)} \right)
(1-q_J)\right) \frac{\theta_{IJ}}{k^2}.
\end{equation}

The perturbed Einstein equations (\ref{Ein1}) and (\ref{Ein2})
lead to the relation
\begin{equation}
\frac{k ^2}{a^2}\Psi =  - 4\pi G \epsilon_m,
\end{equation}
and in the long-wavelength limit $k\rightarrow 0$ we obtain
$\epsilon_m \rightarrow 0$. Hence, the terms with $\epsilon_m $ in
the equation (\ref{pnad_2}) can be neglected  on a large scale.

The symmetrical quantities $\tilde{S}_{IJ}$ are closely related to
the entropy perturbations and adiabatic condition. Using the
definition of the scalar field energy-momentum tensor, the
adiabaticity condition for systems of canonical scalar fields
$\varphi_I$\cite{MFB}
\begin{equation}
\frac{\delta\varphi_I}{\varphi_I'} - \frac{\delta\varphi_J
}{\varphi_J'} = 0
\end{equation}
can be rewritten as
\begin{equation}
\frac{\theta_{IJ}}{k^2} =0,
\end{equation}
where $I,J=1,...,N$.

The adiabatic mode in multi-fluid models with barotropic coupled
fluids is determined by the conditions \cite{Malik_PhD}
\begin{equation}
\hat{\Delta} _{IJ} =0.
\end{equation}

In both cases, quantities $\tilde{S} _{IJ}$ vanish. The case of
systems with barotropic fluids and scalar fields is more
complicated, since the leading contributions to $\tilde{S} _{IJ}$
depend on the gauge choice. In cosmological models with minimally
coupled quintessence dark energy, one can put $\tilde{S} _{IJ}=0$
for the growing adiabatic mode in the longitudinal gauge, but some
of quantities $\tilde{S}_{IJ}$ are different from zero at leading
order in $k\tau$ at "generalized initial adiabatic conditions" of
\cite{1004.5509} in the synchronous gauge. The "generalized
initial adiabatic conditions" provide only that the inequality
$P_{nad} \ll \delta p$ holds. In any case, to find $P_{nad}$ and
the comoving curvature perturbation ${\cal R}$, it is necessary to
know the values of $\tilde{S}_{IJ}$.

The equation (\ref{pnad_2}) shows that the non-adiabatic pressure
perturbation depend on $\hat{\Delta}_{IJ}$ and $\theta_{IJ}$
through their linear combinations $\tilde{S} _{IJ}$. This fact
allows to simplify the analysis of the evolution of the
non-adiabatic pressure perturbation that is important for the
study of possibility of the non-adiabatic early instabilities in
the models with coupled fluids.

The quantities $\tilde{S} _{IJ}$ are completely determined by the
values of $\hat{\Delta}_{IJ}$ and $\theta_{IJ}$. Following Refs.
\cite{Malik_PhD},\cite{Malik_Wands_Ungarelli},\cite{Malik_Wands2004},
we introduce the notations
\begin{equation}
\hat{\epsilon}_I=\frac{\delta Q_I}{\bar{Q}_{I}}+\frac{\bar{Q}_I'}{
3{\cal H} \bar{Q}_I}\Delta, ~~~ f_{IJ} = \frac{af_I}{\rho_I
(1+w_I)} -\frac{af_J}{\rho_J (1+w_J)}, ~~~ \hat{E}_{IJ}
=\frac{q_I\hat{\epsilon}_I}{1-q_I}  - \frac{q_J
\hat{\epsilon}_J}{1-q_J} .
\end{equation}

Then for the system of $N$ coupled fluids (taking into account
$\rho=\sum_I \rho_I$) the equations (\ref{cont_1a}),
(\ref{cont_2a}) yield

\begin{eqnarray}
\hat{\Delta}_{IJ} ' &-& \frac{1}{2}\left(\frac{q_I'}{1-q_I} -3
{\cal H} \left(c^2_{sI} -c^2_{sI(ad)} \right) - 3{\cal H} q_I
\left(c^2_{sI(ad)} +1\right) \right)
\sum_K\frac{\bar{\rho}_K'}{\bar{\rho}'}
\left(\hat{\Delta}_{IK}+ \hat{\Delta}_{JK}\right) \nonumber \\
&&+ \frac{1}{2}\left( \frac{q_J'}{1-q_J} - 3{\cal H} \left(
c^2_{sJ} - c^2_{sJ(ad)} \right) - 3{\cal H} q_J\left(c^2_{sJ(ad)}
+1 \right) \right) \sum_K \frac{\bar{\rho}_K'}{\bar{\rho}'} \left(
\hat{\Delta}_{IK}+ \hat{\Delta}_{JK} \right) \nonumber \\
&&+ \frac{3}{2}{\cal H} \left(c^2_{sI} -c^2_{sI(ad)} + c^2_{sJ} -
c^2_{sJ (ad)}+q_I (c^2_{sI(ad)} +1)+ q_J(c^2_{sJ(ad)}+1)\right)
\hat{\Delta}_{IJ} \nonumber \\
&&- \frac{1}{2}\left(\frac{q_I'}{1-q_I} + \frac{q_J'}{1-q_J}
\right) \hat{\Delta}_{IJ} +\frac{9}{2}{\cal H}^2\left(c^2_{sI} -
c^2_{sI(ad)} + c^2_{sJ(ad)}  - c^2_{sJ} \right)\sum_K
\frac{\bar{\rho}_K +\bar{P}_K}{\bar{\rho}+\bar{P}} \left(\frac{
\theta_{IK}}{k^2} +\frac{\theta_{JK}}{k^2}
\right) \nonumber \\
\label{cont_1b}&&=3{\cal H} \hat{E}_{IJ}  - \left( \frac{1}{ 1-
q_I} -\frac{1 }{1-q_J} \right) \theta_{uc} -3{\cal H} \left(
c^2_{sI} - c^2_{sI(ad)} + c^2_{sJ(ad)}  - c^2_{sJ} \right) \frac{
\epsilon_m }{\bar{\rho}+\bar{P}} ,\\
\frac{\theta_{IJ}'}{k^2} &-& \frac{3}{2}{\cal H}\left(c^2_{sI}
-c^2_{sJ} - q_I(1+c^2_{sI}) + q_J (1+c^2_{sJ})\right) \sum_K
\frac{\bar{\rho}_K+ \bar{P}_K}{\bar{\rho}+\bar{P}}
\left(\frac{\theta_{IK}}{k^2}
+\frac{\theta_{JK}}{k^2}\right) \nonumber \\
&&+ {\cal H} \left(1 -\frac{3}{2}
c^2_{sI} - \frac{3}{2}c^2_{sJ} + \frac{3}{2} q_I(1+c^2_{sI}) +
\frac{3}{2}q_J (1+c^2_{sJ})\right) \frac{\theta_{IJ}}{k^2}  \nonumber \\
&&- \frac{1}{2}\left( c^2_{sI}(1-q_I) -  c^2_{sB}(1-q_J)\right)
\sum_K\frac{\bar{\rho}_K'}{\bar{\rho}'} \left(\hat{\Delta}_{IK} +
\hat{\Delta}_{JK} \right) \nonumber \\
\label{cont_2b}&&- \frac{1}{2} \left(c^2_{sI}(1-q_I)+ c^2_{sJ} (1-
q_J) \right) \hat{\Delta}_{IJ} = \left( (1-q_I) c^2_{sI} -
(1-q_J)c^2_{sJ} \right) \frac{ \epsilon _m}{\bar{\rho}+\bar{P}} -
f_{IJ},
\end{eqnarray}
where $I,J,K=1,...,N$ and
\begin{equation}
\theta_{uc} \equiv \theta + k^2\frac{\psi}{{\cal H}}
\end{equation}
is the velocity perturbation on uniform curvature hypersurfaces.

These equations agree with the equations of Ref. \cite{Malik_PhD}
on a spatially flat background at a vanishing anisotropic stress
\footnote{There is a typo in the common sign of the second line of
Eq. (2.196) of Ref. \cite{Malik_PhD}.}. On a large scale, after
substituting the explicit expressions for $\hat{E}_{AB}$ and
$f_{AB}$, the equations (\ref{cont_1b}),(\ref{cont_2b}) form a
closed system of equations.

The system of equations (\ref{cont_1b}), (\ref{cont_2b}) contains
$N(N-1)$ independent equations for the antisymmetric quantities
$\hat{\Delta}_{IJ}$ and $\theta_{IJ}$. Meanwhile, for the analysis
of the evolution of the non-adiabatic pressure perturbation, it is
suffices to know only the values of $ N (N-1)/2 $ symmetric
quantities $\tilde{S} _{IJ}$. In many cases of practical
importance, on a large scale, it is not need to solve the complete
system of equations (\ref{cont_1b}) and (\ref{cont_2b}), and one
can derive and solve the equations for $\tilde{S} _{IJ}$. In what
follows, we consider some applications of the new variables.

\section{Several models. }
\label{appl}

\subsection{Two non-coupled fluids.}

Models with non-interacting fractions are often used in modern
cosmology, especially in inflationary models. They are of interest
as toy-models when considering the radiation dominated universe.
Primarily, we are interested in studying the influence of
non-adiabatic sound speed on rate of decay of non-adiabatic
perturbations in cosmological models with quintessence. We would
also like to clarify the general arguments \cite{new_approach}
that the non-adiabatic perturbations evolve independently of the
adiabatic ones within the effective fluid formalism applied in the
paper.

\subsubsection{Non coupled barotropic fluid and scalar field.}

Consider a simple model with two minimally coupled perfect fluids,
indicated by subscripts $A$ and $B$,  where the $B$-component is a
barotropic fluid. In this case, the equations (\ref{cont_1b}),
(\ref{cont_2b}) are reduced to
\begin{eqnarray}
\label{min_1}\tilde{\Delta}_{AB}' + 3{\cal H}\left(c^2_{sA}
-c^2_{sA(ad)} \right) \frac{\bar{\rho}_B' }{\bar{\rho}'}
\left(\tilde{\Delta}_{AB}+ 3{\cal H}  \frac{ \theta_{AB}}{k^2}
\right)+ \theta_{AB} &=&- 3{\cal H} \left( c^2_{sA} - c^2_{sA(ad)}
\right)\frac{ \epsilon_m }{\bar{\rho}+\bar{P}}, \\
\label{min_2}\frac{\theta_{AB}'}{k^2} + {\cal H}\frac{ \theta_{AB}
}{k^2}  - \left( c^2_{sA} \frac{\bar{\rho}_B'}{ \bar{\rho}'} +
c^2_{sB} \frac{\bar{\rho}_A '}{\bar{\rho}'} \right)\left( \tilde{
\Delta}_{AB} + 3{\cal H}  \frac{ \theta_{AB}}{k^2} \right) &=&
\left( c^2_{sA} - c^2_{sB}\right) \frac{ \epsilon
_m}{\bar{\rho}+\bar{P}}.
\end{eqnarray}

The  definition (\ref{psevdo}) gives
\begin{equation}
\tilde{S}_{AB}=\left(c^2_{sA} -c^2_{sB}\right) \tilde{\Delta}_{AB}
+3{\cal H} \left(c^2_{sA} - c^2_{sA(ad)}\right) \frac{\theta_{AB}
}{k^2}.
\end{equation}

Here we restrict ourselves only to the case of constant
$c^2_{sA}$, $c^2_{sB}$, $w_A$, $w_B$. At constant parameters
$w_A$, $w_B$ the equation (\ref{constr}) implies the relations
\begin{equation}
c^2_{sA(ad)} = w_A,~~~~~~~~c^2_{sB(ad)} = w_B.
\end{equation}

The equations (\ref{min_1}), (\ref{min_2}) can be combined to form
the second order equation
\begin{eqnarray}
\tilde{S}_{AB}'' &+& {\cal H}\left[1 + 3\left( c^2_{sA} \frac{
\bar{\rho}_B' }{\bar{\rho}'} + c^2_{sB} \frac{\bar{\rho}_A '}{
\bar{\rho}'} \right) \frac{c^2_{sB} - c^2_{sA(ad)} }{c^2_{sA} -
c^2_{sB}} - 3c^2_{sB}\frac{ c^2_{sA} -c^2_{sA(ad)} }{c^2_{sA}
-c^2_{sB} } - \frac{\xi '}{{\cal H}\xi} \right]
\tilde{S}_{AB}' \nonumber \\
&&+3{\cal H}^2 \left[\left( c^2_{sA} \frac{\bar{\rho}_B'}{
\bar{\rho}'} + c^2_{sB} \frac{\bar{\rho}_A '}{\bar{\rho}'} \right)
\left(1 - \frac{{\cal H}'}{{\cal H}^2} \right) - \left(1 +
\frac{{\cal H}'}{{\cal H}^2} - \frac{\xi'}{{\cal H}\xi} \right)
c^2_{sB}\right]\frac{ c^2_{sA}- c^2_{sA(ad)} }{c^2_{sA} -c^2_{sB} }
\tilde{S}_{AB} \nonumber \\
\label{diff_entr_min}&&+\left( c^2_{sA} \frac{\bar{\rho}_B'
}{\bar{\rho}'} + c^2_{sB} \frac{\bar{\rho}_A '}{\bar{\rho}'}
\right) k^2\tilde{S}_{AB} = \xi\left( c^2_{sA} - c^2_{sB}  \right)
\frac{ \epsilon _m}{\bar{\rho}+\bar{P}},
\end{eqnarray}
where
\begin{equation}
\xi= -\left[\left(c^2_{sA} -c^2_{sB} \right) k^2+ 3{\cal H}^2
\left(1 - \frac{{\cal H}'}{{\cal H}^2}+ 3 c^2_{sB}\frac{c^2_{sB}
-c^2_{sA(ad)} }{c^2_{sA} - c^2_{sB} } \right) \left(c^2_{sA} -
c^2_{sA(ad)}\right)\right].
\end{equation}

For two barotropic fluids with the indices $A$ and $B$, the
equation (\ref{diff_entr_min}) is reduced to
\begin{equation}
\label{diff_entr1}\tilde{S}_{AB}'' + {\cal H}\left[1 - 3\left(
c^2_{sA} \frac{\bar{\rho}_B' }{\bar{\rho}'} + c^2_{sB}
\frac{\bar{\rho}_A '}{\bar{\rho}'} \right) \right] \tilde{S}_{AB}'
+ k^2 \left( c^2_{sA} \frac{\bar{\rho}_B' }{\bar{\rho}'} + c^2_{sB}
\frac{\bar{\rho}_A '}{\bar{\rho}'} \right) \tilde{S}_{AB} =-
k^2\left( c^2_{sA} - c^2_{sB} \right)^2 \frac{ \epsilon
_m}{\bar{\rho}+\bar{P}}.
\end{equation}
The additional multiplier $k^2$ on the right hand side of
(\ref{diff_entr1}) ensures that in the large scale limit
$k\rightarrow 0$, the source term is negligibly small, and the
adiabatic perturbations do not affect the entropy evolution.

When the $A$-component is a canonical scalar field, the implicit
form of the equation (\ref{diff_entr_min}) looks rather
complicated. We write it only for the case of radiation and
subdominant scalar field. Then, on a large scale, we obtain
\begin{equation}
\label{diff_entr2}\tilde{S}_{AB}'' + 3{\cal H}\left(1 -
c^2_{sA(ad)}\right) \tilde{S}_{AB}' +6{\cal H}^2(1 -c^2_{sA(ad)} )
\tilde{S}_{AB} =- {\cal H}^2 \left(5 -3c^2_{sA(ad)} \right)
\left(1 - c^2_{sA(ad)}\right) \frac{ \epsilon
_m}{\bar{\rho}+\bar{P}}.
\end{equation}

Although this equation is gauge invariant, according to the
definition of $\tilde{S}_{AB}$, one can drop the source only if
$max\{|\Delta_A|,|\Delta_B|, {\cal H} |\theta_A/k^2|,{\cal H}
|\theta_B/k^2|\} \gg \frac{ \epsilon _m}{\rho+p}$. This condition
is not satisfied, in general, in the synchronous gauge.

Indeed, for the "generalized adiabatic initial conditions" of Ref.
\cite{1004.5509}
\begin{eqnarray}
\Delta_A &=&-\frac{C}{2}\frac{4-3c_{sA}^2}{4 -6w_A+3c_{sA}^2}
(k\tau)^2,\\
\theta_A &=&-\frac{C}{2} \frac{c_{sA}^2}{4-6w_A+3c_{sA}^2}
(k\tau)^3k,
\end{eqnarray}
we get
\begin{equation}
\tilde{S}_{AB}= \left(1- w_A \right) \frac{10-6w_A}{3(7-6w_A)} \frac{C}{2}
(k\tau)^2, ~~~~~~~~ \frac{\epsilon_m}{\bar{\rho} +\bar{P}} = -
\frac{4}{3}\frac{C}{2} (k\tau)^2,
\end{equation}
where $C$ is a constant. A direct verification shows that
(\ref{diff_entr2}) holds for the "generalized adiabatic initial
conditions" only if we do not neglect the source in this equation.
Thus, in the synchronous gauge, the adiabatic perturbations give
rise to non-adiabatic perturbations even on a large scale.

It is interesting to compare equation (\ref{diff_entr1}) and
(\ref{diff_entr2}) at the radiation dominated stage in case the
quantities $\tilde{S}_{AB}$ are large enough for the sources to be
neglected. The equation (\ref{diff_entr1}) yields on a large scale
\begin{equation}
\label{diff_entr1b}\tilde{S}_{AB}'' + {\cal H}\left(1 - 3 w_{A}
\right) \tilde{S}_{AB}'=0 ~~~~~(\mathrm{fluid - fluid}).
\end{equation}
The equation (\ref{diff_entr2}) takes the form
\begin{equation}
\label{diff_entr2b}\tilde{S}_{AB}'' + {\cal H}\left(3 -
3w_{A}\right) \tilde{S}_{AB}' +6{\cal H}^2(1 -w_{A} )
\tilde{S}_{AB} = 0 ~~~~~(\mathrm{scalar~ field - fluid}).
\end{equation}
It follows that, if the $A$-fluid is the minimally coupled scalar
field, the damping force is stronger. In addition, when $ w_A <1
$, the coefficient at $\tilde{S}_{AB}$ is positive. Hence, at
negative $w_A$, the quantity $\tilde{S}_{AB}$ rapidly approaches
the asymptotic solution. For this reason, in numerical codes like,
CMBFAST and CAMB, the initial dark energy perturbations are set by
default to zero.

\subsubsection{Two scalar fields.}

The case of two non-interacting scalar fields, and the more
general case of interacting fields are studied carefully by
different methods. Here, we look at the second order equation for
$\tilde{S}_{AB}$ only. From the original equations (\ref{cont_1b})
and (\ref{cont_2b}), it is easy to write an equation for
$\theta_{AB}$. It has the form
\begin{eqnarray}
\frac{\theta_{AB}''}{k^2} &+& {\cal H}\left( 1 - 3\left( c^2_
{sA(ad)} \frac{\bar{\rho}_B' }{ \bar{\rho}'} + c^2_{sB(ad)}
\frac{\bar{\rho}_A' }{ \bar{\rho}'}\right) \right)
\frac{\theta_{AB}'}{k^2} + \theta_{AB}
\nonumber \\
&&+ {\cal H}^2 \left( 3 - 2\frac{{\cal H}' }{{\cal H}^2}- 3\left(
c^2_{sA(ad)} \frac{\bar{\rho}_B' }{ \bar{\rho}'} +  c^2_{sB(ad)}
\frac{\bar{\rho}_A' }{ \bar{\rho}'}\right) \right) \frac{
\theta_{AB}}{k^2} = 3{\cal H} \left(  c^2_{sA(ad)} - c^2_{sB(ad)}
\right) \frac{ \epsilon_m}{\bar{\rho}+\bar{P}} .
\end{eqnarray}

At constant adiabatic sound speeds, this equation can be rewritten
in terms of $\tilde{S}_{AB}$ as
\begin{eqnarray}
\tilde{S}_{AB}'' &+& {\cal H}\left( 1 - 2\frac{{\cal H}'}{{\cal
H}} - 3\left( c^2_{sA(ad)} \frac{\bar{\rho}_B' }{ \bar{\rho}'} +
c^2_{sB(ad)} \frac{\bar{\rho}_A'}{ \bar{\rho}'}\right) \right)S'
+k^2\tilde{S}_{AB} \nonumber  \\
&&+ {\cal H}^2 \left( 2 \frac{{{\cal H}'}^2}{{\cal H}^4} -
\frac{{\cal H}'' }{{\cal H}^3} + 3\left(1  -  c^2_{sA(ad)}
\frac{\bar{\rho}_B' }{ \bar{\rho}'} - c^2_{sB(ad)}
\frac{\bar{\rho}_A' }{\bar{ \rho}'} \right) \left(1 - \frac{{\cal
H}'}{{\cal H}^2}\right)\right)\tilde{S}_{AB} \nonumber \\
\label{diff_entr_sc}&&=- 9{\cal H}^2 \left( c^2_{sA(ad)} -
c^2_{sB(ad)} \right)^2 \frac{ \epsilon_m}{\bar{\rho}+\bar{P}} .
\end{eqnarray}

In general, one has to take into account the source term even on a
large scale, but usually the scalar fields are considered in the
longitudinal ($E=B=0$) gauge. The well known long-wavelength
adiabatic solution in this gauge \cite{PSt92,PSt94} can be
rewritten as
\begin{eqnarray}
\phi &=& \psi = C\left( 1 - \frac{{\cal H}}{a^2} \int_{\tau_1}
^{\tau} a^2d\tau\right), \\
\Delta_A &=& \Delta_B=-3 C  \frac{{\cal H}}{a^2}\int_{\tau_1}^{\tau} a^2d\tau,\\
\frac{\theta_A}{k^2} &=& \frac{\theta_B}{k^2} =\frac{C}{a^2}
\int_{\tau_1}^{\tau} a^2d\tau,
\end{eqnarray}
where $C$ and $\tau_1$ are constants. By the definition
(\ref{epsilon_m}), the value of $\epsilon_m$ vanishes now in the
leading order, and the right hand side of the equation
(\ref{diff_entr_sc}) can be neglected.

\subsection{Special case of three fluids.}

Consider the Universe filled with quintessence dark energy ($A$),
dark matter ($B$) and radiation ($C$), and assume that the dark
energy and the dark matter couple. Since the variables
$\hat{\Delta}_{IJ}$, $\theta_{IJ}$ are antisymmetric and
constrained by
\begin{equation}
\label{cyclic}\hat{\Delta}_{AB}+\hat{\Delta}_{BC}+\hat{\Delta}_{CA}=0,
~~~~~~ \theta_{AB}+\theta_{BC}+\theta_{CA}=0,
\end{equation}
there are only four independent variables among them.

Using the relations (\ref{cyclic}) and the equalities
$c^2_{sB}=c^2_{sB(ad)}$, $c^2_{sC}=c^2_{sC(ad)}$, which are valid
for barotropic fluids, the equations (\ref{cont_1b}) and
(\ref{cont_2b}) can be reduced to set
\begin{eqnarray}
\hat{\Delta}_{AC} ' &-& \left(\frac{q_A'}{1-q_A}-3{\cal H} \left(
c^2_{sA} -c^2_{sA(ad)} \right) - 3{\cal H} q_A (c^2_{sA(ad)} +1)
\right) \left( \frac{\bar{\rho}_B'+\bar{\rho}_C'}{\bar{\rho}'}
\hat{\Delta}_{AC}-\frac{\bar{\rho}_B'}{\bar{\rho}'}
\hat{\Delta}_{BC} \right) \nonumber \\
&&+9{\cal H}^2\left(c^2_{sA} - c^2_{sA(ad)} \right) \left( \frac{
\bar{\rho}_B +\bar{P}_B+\bar{\rho}_C +\bar{P}_C }{\bar{\rho} +
\bar{P}} \frac{ \theta_{AC}}{k^2} - \frac{ \bar{\rho}_B + \bar{P}
_B}{\bar{\rho}+\bar{P}}\frac{ \theta_{BC}}{k^2} \right) \nonumber \\
&&+ \left(\frac{ \bar{\rho}_B +\bar{P}_B+\bar{\rho}_C +\bar{P}_C
}{\bar{\rho} + \bar{P}} \frac{1 }{1 - q_A} + \frac{\bar{\rho}_A +
\bar{P}_A}{\bar{\rho}+\bar{P}} \right) \theta_{AC}-\frac{q_A
}{1-q_A} \frac{\bar{\rho}_B + \bar{P}_B}{\bar{\rho}+\bar{P}}
\theta_{BC} \nonumber \\
&&=3{\cal H} \hat{E}_{AC}  - \frac{ q_A }{1-q_A} \theta_{uc}
-3{\cal H}\left(c^2_{sA} -c^2_{sA(ad)}  \right) \frac{
\epsilon_m}{ \bar{\rho}+\bar{P}} , \\
\frac{\theta_{AC}'}{k^2} &+& {\cal H} \left(1 - 3c^2_{sC} \right)
\frac{\theta_{AC}}{k^2} - 3{\cal H}\left( c^2_{sA} -c^2_{sC}  -
q_A(1+c^2_{sA}) \right) \left( \frac{ \bar{\rho}_B +\bar{P}_B
+\bar{\rho}_C +\bar{P}_C }{\bar{\rho} + \bar{P}} \frac{\theta_{AC}
}{k^2} - \frac{\bar{\rho}_B+ \bar{P}_B}{ \bar{\rho}+\bar{P}}
\frac{\theta_{BC}}{k^2}\right) \nonumber \\
&& - \left((1-q_A) c^2_{sA} -  c^2_{sC}\right) \left( \frac{
\bar{\rho}_B'+ \bar{\rho}_C' }{\bar{\rho}'} \hat{\Delta}_{AC} -
\frac{\bar{\rho}_B' }{\bar{\rho}'} \hat{\Delta} _{BC} \right) -
c^2_{sC} \hat{\Delta}_{AC} = \left((1-q_A) c^2_{sA} - c^2_{sC}
\right) \frac{ \epsilon_m}{\bar{\rho}+\bar{P}} - f_{AC}, \\
\hat{\Delta}_{BC} ' &-& \left(\frac{q_B'}{1-q_B} - 3{\cal H} q_B
(c^2_{sB(ad)} +1) \right) \left( \frac{\bar{\rho}_A' + \bar{\rho}
_C'}{\bar{\rho}'} \hat{\Delta}_{BC} - \frac{\bar{\rho}_A'
}{\bar{\rho}'}\hat{\Delta}_{AC}\right) \nonumber \\
&&+ \left( \frac{\bar{\rho}_B+ \bar{P}_B}{\rho+p} + \frac{
\bar{\rho}_A+ \bar{P}_A+\bar{\rho}_C+ \bar{P}_C}{ \bar{\rho}
+\bar{P}} \frac{1 }{1-q_B}\right) \theta_{BC}-\frac{\bar{\rho}_A+
\bar{P}_A}{\bar{\rho}+\bar{P}} \frac{q_B }{1-q_B} \theta_{AC} =
3{\cal H}\hat{E}_{BC} - \frac{q_B}{1-q_B} \theta_{uc} ,\\
\frac{\theta_{BC}'}{k^2} &+& {\cal H} \left(1 - 3 c^2_{sC} \right)
\frac{\theta_{BC}}{k^2} - 3{\cal H}\left( c^2_{sB} -c^2_{sC} -
q_B(1 + c^2_{sB}) \right)\left(\frac{\bar{\rho}_A+ \bar{P}_A +
\bar{\rho}_C+ \bar{P}_C}{\bar{\rho}+\bar{P}} \frac{\theta_{BC}}
{k^2} - \frac{\bar{\rho}_A+ \bar{P}_A}{\bar{\rho}+\bar{P}}
\frac{\theta_{AC}}{k^2}\right) \nonumber \\
&&- \left((1-q_B) c^2_{sB} -  c^2_{sC}\right)\left( \frac{ \bar{
\rho}_A' + \bar{\rho}_C' }{\bar{\rho}'} \hat{\Delta}_{BC} - \frac{
\bar{\rho}_A' }{\bar{\rho}'} \hat{\Delta }_{AC} \right) - c^2_{sC}
\hat{\Delta}_{BC} = \left((1-q_B) c^2_{sB} - c^2_{sC}\right)
\frac{ \epsilon _m}{\bar{\rho}+\bar{P}} - f_{BC}.
\end{eqnarray}

At the radiation dominated stage the dark energy and dark matter
are subdominant, $\frac{\rho_A}{\rho}\ll 1$,
$\frac{\rho_B}{\rho}\ll 1$, which allows to simplify the
equations. Furthermore, we assume that we can neglect terms with
$\frac{ \epsilon _m}{\rho+p}$ in the long-wavelength limit. As a
result, on a large scale, we obtain the approximate equations
\begin{eqnarray}
\label{appox1}\tilde{\Delta}_{AC} ' + 3{\cal H}\left(  c^2_{sA}
-c^2_{sA(ad)} + q_A (c^2_{sA(ad)} +1) \right) \tilde{\Delta}_{AC}
+ 9{\cal H}^2 (c^2_{sA} - c^2_{sA(ad)})(1-q_A)\frac{
\theta_{AC}}{k^2} &=& 3{\cal H}(1-q_A)
\hat{E}_{AC} , \\
\label{appox2}\frac{\theta_{AC}'}{k^2} + {\cal H}\left(1 -
3c^2_{sA}  + 3q_A(1+c^2_{sA}) \right) \frac{\theta_{AC}}{k^2} -
c^2_{sA}\tilde{\Delta}_{AC} &=&   - f_{AC}, \\
\label{appox3}\tilde{\Delta}_{BC} ' +
3{\cal H} q_B (c^2_{sB(ad)} +1)  \tilde{\Delta}_{BC} +
\theta_{BC}+ q_B \theta_{uc}  &=& 3{\cal H}(1-q_B) \hat{E}_{BC} , \\
\label{appox4}\frac{\theta_{BC}'}{k^2} + {\cal H} \left(1 - 3
c^2_{sB} +3q_B(1 + c^2_{sB}) \right) \frac{\theta_{BC}}{k^2} -
c^2_{sB}\tilde{\Delta}_{BC} &=& - f_{BC}.
\end{eqnarray}
In the following we explicitly use the values $c^2_{sA}=1$,
$c^2_{sB}=0$, $c^2_{sC}=1/3$.

\subsubsection{Example 1.}

As a simple model, consider the coupling \cite{0912.0120}
\begin{equation}
Q^\mu =\gamma \rho_A\rho_B (u^\mu_B-u^\mu_{A}),
\end{equation}
where $\gamma$  is a constant.

For such interaction, all the quantities $\bar{Q}_A$ are zero and
the evolution of background variables is the same as at minimally
coupled fractions. Now $ q_A=q_B = \epsilon_A =\epsilon_B =0$ and
\begin{equation}
f_{AC} \approx \frac{a\gamma \bar{\rho}_B}{1+w_A} \frac{\theta_{AB}
}{k^2}, ~~~~~~~ f_{BC} \approx -a\gamma \bar{\rho}_A
\frac{\theta_{AB}}{k^2}.
\end{equation}

The equations (\ref{appox3}) and (\ref{appox4}) give
\begin{eqnarray}
\label{appox3_a}\hat{\Delta}_{BC} ' + \theta_{BC} &=& 0 , \\
\label{appox4_a}\frac{\theta_{BC}'}{k^2} + {\cal H}
\frac{\theta_{BC}}{k^2} &=& a\gamma \bar{\rho}_A
\frac{\theta_{AB}}{k^2}.
\end{eqnarray}

We consider here only the case of $w_A<-1/3$. Since for minimally
coupled fractions $\bar{\rho}_A\propto a^{-3(1+w_A)}$ and $\left|
\frac{a\gamma \bar{\rho}_A}{{\cal H}}\right| \ll 1$ at early
times, we find that one can consistently assume that
$\tilde{\Delta}_{BC}= \theta_{BC} =0$ if $\theta_{AB}$ are not
very large. Then the first two equations, (\ref{appox1}) and
(\ref{appox2}), are simplified to
\begin{eqnarray}
\tilde{\Delta}_{AC} ' + 3{\cal H} \left( 1 -w_{A} \right)
\tilde{\Delta}_{AC} + 9{\cal H}^2 (1 -w_{A} )
\frac{ \theta_{AC}}{k^2} &=& 0 , \\
\frac{\theta_{AC}'}{k^2} + {\cal H}\left(\frac{a\gamma
\bar{\rho}_B}{{\cal H}(1+w_A)} - 2 \right) \frac{\theta_{AC}}{k^2}
- \tilde{\Delta}_{AC} &=& 0.
\end{eqnarray}

These equations can be rewritten in the equivalent form
\begin{eqnarray}
\tilde{S}_{AC}' + {\cal H} \left( 1+ \frac{a\gamma \bar{\rho}_B
}{{\cal H} (1+w_A)} \right) \tilde{S}_{AC} &=& {\cal H} \left[
\frac{5}{3} -w_{A}  + \frac{2}{3} \frac{a\gamma \bar{\rho}_B}{
{\cal H} (1+w_A)}= \right] \tilde{ \Delta} _{AC}, \\
\tilde{\Delta}_{AC} ' + {\cal H} \left( 1 -3w_{A} \right)
\tilde{\Delta}_{AC} &=& - 3{\cal H}\tilde{S}_{AC} .
\end{eqnarray}

Concerning the sufficiently early Universe, we have
$\left|\frac{a\gamma \bar{\rho}_B}{{\cal H} (1+w_A)} \right|\gg 1$.
Under this condition one can obtain the second order equation
\begin{equation}
\tilde{S}_{AC}'' + {\cal H} \frac{a\gamma \bar{\rho}_B}{ {\cal H}
(1+w_A)} \tilde{S}_{AC}' +3{\cal H} ^2 \left( 1 - w_{A} \right)
\frac{a\gamma \bar{\rho}_B}{{\cal H} (1+w_A)}  \tilde{S}_{AC} =0.
\end{equation}

The presence or absence of instabilities is completely determined
by the sign of the coupling constant $\gamma$. At negative
$\gamma$, there are an anti-damping force and tachyonic
instability, which increase indefinitely in the limit
$a\rightarrow 0$ and lead to a catastrophic growth of
$\tilde{S}_{AC}$ and a non-adiabatic pressure perturbation
$P_{nad}$.

\subsubsection{Example 2.}

Another simple coupling is $Q_A=-Q_B=\gamma \rho_B$
\cite{VMM_0804}. At the radiation dominated stage the background
dark sector densities are
\begin{equation}
\label{dark} \bar{\rho}_A = \frac{a \gamma
\tau}{3w_A+2}\bar{\rho}_B,~~~~~ \bar{\rho}_B \propto a^{-3}.
\end{equation}
This background solution yields
\begin{equation}
\tilde{E}_{AC} \approx q_A \tilde{\Delta} _{BC}, ~~~~~~
\tilde{E}_{BC} \approx 0,
\end{equation}
\begin{equation}
f_{AC} = -\frac{1}{\tau} \frac{3w_A+2}{1+w_A}
\frac{\theta_{BC}}{k^2}, ~~~~~~ f_{BC} \approx 0,
\end{equation}
where
\begin{equation}
\label{coupl}q_{A}=\frac{3w_A+2}{3(1+w_A)}.
\end{equation}

The equations (\ref{appox1})-(\ref{appox4}) reduce now to
\begin{eqnarray}
\label{approx1}\tilde{\Delta}_{AC} ' + 3{\cal H}\left( 1 -w_{A} +
q_A (w_{A} +1) \right) \tilde{\Delta}_{AC} + 9{\cal H}^2 (1 -
w_{A}) (1-q_A)\frac{ \theta_{AC}}{k^2} &=& 3{\cal H} q_A
\tilde{\Delta}
_{BC} , \\
\label{approx2}\frac{\theta_{AC}'}{k^2} + {\cal H}\left(6q_A  -
2\right) \frac{\theta_{AC}}{k^2} - \tilde{\Delta}_{AC} &=&
\frac{1}{\tau} \frac{3w_A+2}{1+w_A} \frac{\theta_{BC}}{k^2}
\end{eqnarray}
and
\begin{eqnarray}
\tilde{\Delta}_{BC} ' + \theta_{BC} &=& 0, \\
\frac{\theta_{BC}'}{k^2} + {\cal H}\frac{\theta_{BC}}{k^2}  &=& 0.
\end{eqnarray}

From the last two equations it follows that one can consistently
assume that $\tilde{\Delta}_{BC}= \theta_{BC}=0$, if this
condition was fulfilled initially. The first two equations
(\ref{approx1}) and (\ref{approx2}) then give
\begin{eqnarray}
\tilde{\Delta}_{AC} ' + 3{\cal H}\left( 1 -w_{A} + q_A (w_{A} +1)
\right) \tilde{\Delta}_{AC} + 9{\cal H}^2 (1 - w_{A})
(1-q_A)\frac{ \theta_{AC}}{k^2} &=& 0,\\
\frac{\theta_{AC}'}{k^2} + {\cal H}\left( 6q_A - 2 \right)
\frac{\theta_{AC}}{k^2} &=& \tilde{\Delta}_{AC} .
\end{eqnarray}

Using the definition of $\tilde{S}_{AC}$, the equation
(\ref{coupl}), and substituting ${\cal H}=1/\tau$, one can write
now the second order equation
\begin{equation}
\tilde{S}_{AC}''+ \frac{1}{\tau} \frac{11w_A+9}{1+w_A}
\tilde{S}_{AC}' + \frac{2}{\tau^2} \frac{11w_A+9}{1+w_A}
\tilde{S}_{AC} =0 .
\end{equation}
Assuming a power-law form of the solution $\tilde{S}_{AC}\propto
\tau^{\tilde{n}}$,  we obtain the algebraic equation for the
power-law index
\begin{equation}
\tilde{n}^2+ \frac{10w_A+8}{1+w_A}\tilde{n}  + 2
\frac{11w_A+9}{1+w_A} =0 ,
\end{equation}
with the roots
\begin{equation}
\tilde{n}_\pm=-\frac{5w_A+4}{1+w_A} \pm \frac{\sqrt{3w_A^2 -
2}}{1+w_A}.
\end{equation}
The equations (\ref{pnad_2}) and (\ref{dark}) imply that on a
large scale $P_{nad}\approx\frac{\rho_A}{\rho} \tilde{S}_{AC}
\propto \tau^3\tilde{S}_{AC}$. Hence, $P_{nad}= C_1\tau^{n_+} +
C_2\tau^{n_-}$, where $ C_1 $, $ C_2 $ are constants and
\begin{equation}
n_\pm=-\frac{2w_A+1}{1+w_A} \pm \frac{\sqrt{3w_A^2 - 2}}{1+w_A}.
\end{equation}
The resulting expression agrees with the equation (73) of
\cite{VMM_0804} and a further analysis is identical to the one
given in Ref. \cite{VMM_0804}.

\section{Conclusion. }
\label{concl}

In this paper we considered the evolution of the non-adiabatic
perturbations in the models with multiple interacting fluids with
$c_{sI}^2 \neq c_{sI(ad)}^2$. We wrote the gauge invariant
equations for the variables that determine the non-adiabatic
pressure perturbation and the rate of changes of the comoving
curvature perturbation ${\cal R}$ in the models with energy
exchange . The analysis of these equations was made for several
particular models. One can see that the quantities
$\tilde{S}_{IJ}$, introduced in the paper, allow to clarify the
features of the non-adiabatic perturbations evolution in
multi-fluid cosmological models.




\begin{thebibliography}{99}

\bibitem{Kodama_Sasaki}
H. Kodama and M. Sasaki, Prog. Theor. Phys. Suppl. \textbf{78}, 1
(1984).

\bibitem{LL93}
A. R. Liddle and D. H. Lyth, Phys. Rep. \textbf{231}, 1 (1993).

\bibitem{Linde_1985}
A. D. Linde, Phys. Lett. \textbf{B158}, 375 (1985).

\bibitem{Starobinsky_1985}
A. A. Starobinsky, JETP Lett. \textbf{42}, 152 (1985).

\bibitem{KLF_1985}
L. A. Kofman, A. D. Linde, A. A. Starobinsky, Phys. Lett.
\textbf{B157}, 361, (1985).


\bibitem{KLS_preheat94}
L. Kofman, A. D. Linde and A. A. Starobinsky, Phys. Rev. Lett.
\textbf{73}, 3195 (1994).

\bibitem{KLS_preheat97}
L. Kofman, A. D. Linde and A. A. Starobinsky, Phys. Rev.
\textbf{D56}, 3258 (1997).

\bibitem{LSWL}
S. M. Leach, M. Sasaki, D. Wands, and A. R. Liddle, Phys. Rev.
\textbf{D64}, 023512 (2001).

\bibitem{0806.1016}
L. Boubekeur, P. Creminelli, J. Norena and F. Vernizzi, JCAP
\textbf{0808}, 028 (2008).

\bibitem{Gordon_Hu}
C. Gordon and W. Hu, Phys.Rev. \textbf{D70}, 083003 (2004).

\bibitem{Malik_Wands2004}
K. A. Malik and D. Wands, JCAP \textbf{0502}, 007 (2005).

\bibitem{VMM_0804}
J. Valiviita, E. Majerotto  and R. Maartens, JCAP \textbf{0807},
020 (2008).

\bibitem{0807.3471}
J.-H. He, B. Wang, E. Abdalla, Phys. Lett. \textbf{B671}, 139
(2009).

\bibitem{0901.1611}
M. B. Gavela, D. Hernandez,  L. Lopez Honorez, O. Mena and S.
Rigolin, JCAP \textbf{0907}, 034 (2009).

\bibitem{0901.3272}
B. M. Jackson, A. Taylor, A. Berera, Phys. Rev. \textbf{D79},
043526 (2009).

\bibitem{PSt92}
D. Polarski, A. A. Starobinsky, Nucl. Phys. \textbf{B385}, 623
(1992).

\bibitem{PSt94}
D. Polarski, A. A. Starobinsky, Phys. Rev. \textbf {D50}, 6123
(1994).

\bibitem{GWBM}
C. Gordon, D. Wands, B. A. Bassett and R. Maartens, Phys. Rev.
\textbf{D63} 023506 (2001).

\bibitem{Hwang_Noh}
J. Hwang, H. Noh, Phys.Lett. \textbf{B495}, 277 (2000).

\bibitem{Malik_Wands_Ungarelli}
K. A. Malik, D. Wands and C. Ungarelli, Phys. Rev. \textbf{D67},
063516 (2003).

\bibitem{1004.5509}
G. Ballesteros and J. Lesgourgues, JCAP \textbf{1010}, 014 (2010).

\bibitem{MFB}
V. F. Mukhanov, H. A. Feldman and R. H. Brandenberger, Phys. Rep.
\textbf{215}, 203 (1992).

\bibitem{Bardeen}
J. M. Bardeen, Phys. Rev. \textbf{D22}, 1882 (1980).

\bibitem{Malik_PhD}
K. A. Malik,  arXiv:astro-ph/0101563 .

\bibitem{new_approach}
D. Wands, K. A. Malik, D. H. Lyth and A. R. Liddle, Phys. Rev.
\textbf{D62}, 043527 (2000).

\bibitem{0912.0120}
N. A. Koshelev, arXiv:0912.0120 [gr-qc].


\end{thebibliography}
\end{document}